\documentclass[sigconf]{acmart}
\usepackage{array}

\graphicspath{{./eps/}}
\usepackage{subfigure} 

\renewcommand\footnotetextcopyrightpermission[1]{} 

\begin{document}

\title{A Variable Vector Length SIMD Architecture for\\HW/SW Co-designed Processors}
\titlenote{This work was done while R.Kumar was at UPC Barcelona, A. Mart\'{\i}nez was at Intel Labs, Barcelona and A. Gonz\'{a}lez was at UPC Barcelona and Intel Labs, Barcelona.}

\author{Rakesh Kumar}
\affiliation{%
  \institution{NTNU, Norway}
  \country{}
}

\author{Alejandro Mart\'{\i}nez}
\affiliation{%
  \institution{ARM, UK}
  \country{}
}

\author{Antonio Gonz\'{a}lez}
\affiliation{%
  \institution{UPC Barcelona, spain}
  \country{}
}

\begin{abstract}
Hardware/Software (HW/SW) co-designed processors provide a promising solution to the power and complexity problems of the modern microprocessors by keeping their hardware simple. Moreover, they employ several runtime optimizations to improve the performance. One of the most potent optimizations, vectorization, has been utilized by modern microprocessors, to exploit the data level parallelism through SIMD accelerators. Due to their hardware simplicity, these accelerators have evolved in terms of width from 64-bit vectors in Intel MMX to 512-bit wide vector units in Intel Xeon Phi and AVX-512. Although SIMD accelerators are simple in terms of hardware design, code generation for them has always been a challenge. Moreover, increasing vector lengths with each new generation add to this complexity. 
This paper explores the scalability of SIMD accelerators from the code generation point of view. We discover that the SIMD accelerators remain underutilized at higher vector lengths mainly due to: a) reduced dynamic instruction stream coverage for vectorization and b) increase in permutations. Both of these factors can be attributed to the rigidness of the SIMD architecture. We propose a novel SIMD architecture that possesses the flexibility needed to support higher vector lengths. Furthermore, we propose Variable Length Vectorization and Selective Writing in a HW/SW co-designed environment to transparently target the flexibility of the proposed architecture. We evaluate our proposals using a set of SPECFP2006 and Physicsbench applications. Our experimental results show an average dynamic instruction reduction of 31\% and 40\% and an average speed up of 13\% and 10\% for SPECFP2006 and Physicsbench respectively, for 512-bit vector length, over the scalar baseline code.
\end{abstract}

\settopmatter{printfolios=false}
\maketitle
\pagestyle{plain} 

\section{Introduction}
\label{sec:introduction}
Hardware/Software (HW/SW) co-designed processors offer a solution to the power and complexity problems of modern microprocessors \cite{Sathaye99boa:targeting}\cite{Dehnert:2003:TCM:776261.776263}\cite{Krewell2003}. In order to reduce the power consumption and complexity, these processors incorporate simple hardware. Moreover, several dynamic optimizations are applied to improve the performance. 

Single Instruction Multiple Data (SIMD) accelerators form an integral part of modern microprocessors. Since these accelerators perform the same operation on multiple pieces of data, they just require duplicated functional units and a very simple control mechanism. Despite their simplicity, they are well suited to exploit data level parallelism from modern multimedia, scientific and throughput computing applications. For this reason, SIMD accelerators are ubiquitous in processors from different computing domains like general purpose processors \cite{IntelSWDMan}\cite{848475}\cite{Lee:1996:SPM:623270.624025}, Digital Signal Processors \cite{Arcy1999}, gaming consoles \cite{Kahle2005}\cite{Sporny2002} as well as embedded architectures \cite{Baron2005}. Due to their hardware simplicity, SIMD accelerators grow in size with each new generation. For example, Intel MMX \cite{IntelSWDMan} had vector length of 64-bits, which was increased to 128-bits in SSE \cite{IntelSWDMan} extensions. Intel AVX \cite{IntelSWDMan} and AVX2 \cite{IntelSWDMan} support 256-bit vectors. Whereas Intel`s recent SIMD extensions AVX-512 \cite{IntelAVX512} and Many Integrated Core architecture \cite{IntelMIC} support 512-bit vector operations.
 
In spite of their hardware simplicity, code generation for SIMD accelerators has always been challenging. In the early days, programmers used to target these extensions mainly using in-line assembly or specialized library calls which is tedious and error prone. Then, automatic generation of SIMD instructions (auto-vectorization) was introduced in compilers \cite{Bik:2002:AIV:586554.586555}\cite{Naishlos2004}, which borrowed their methodology from vector compilers. These compilers target loops for generating code for SIMD accelerators. Later, S. Larsen et al. \cite{Larsen:2000:ESL:349299.349320} introduced Superword Level Parallelism (SLP) in which they target basic blocks instead of whole loops for vectorization. Apart from these static approaches, dynamic vectorization in superscalar processors has also been explored by A. Pajuelo et al. \cite{1003585}.

Although SIMD accelerators are amenable to scaling from the hardware point of view, generating efficient code for higher vector lengths is not straightforward. The problem lies in the fact that different applications have different natural vector length. There are applications for which compilers just need to unroll loops with a higher unroll factor to fill the wider vector paths. However, there are other applications that do not have enough parallelism for vectorization at higher vector lengths and SIMD resources are left un/under-utilized. Generating code for these applications for wider vector units becomes a challenge. 

In this paper, we explore the scalability of SIMD accelerators from the code generation point of view. We discover that there are two key factors that thwart the performance at higher vector lengths. First, the dynamic instruction stream coverage for vectorization reduces as vector length increases. This is because the instructions in current vector ISAs operate on all the vector lanes together and not on a subset of it. For example, ADDPS in Intel SSE, VADD in ARM Neon and VADDFP in PowerPC Altivec all operate on all the vector lanes together. Therefore, compilers generate a vector instruction only when there are sufficient numbers of independent operations to fill the vector path. When there are not enough instructions to fill up the vector path, all the instructions are left in the scalar form. We propose to have a flexible SIMD architecture that allows to operate on any number of vector lanes. In addition, we propose Variable Length Vectorization (VLV) to target the flexible vector datapath. 

Second, the number of permutation instructions increases with vector length. The rigidness of SIMD architecture is again responsible for this. For example, the scalar SIMD instructions in Intel SSE always write their result to the lowest element of the vector register. If a vector instruction needs to read these results, they first need to be packed together in a single vector register using shuffle instructions. The proposed SIMD architecture allows scalar instructions to write their result to any element of the vector register depending on how they are needed by the consumer vector instruction. Therefore, the shuffle instructions are no longer required. We call this ability of writing to any selective part of a vector register as Selective Writing (SWR). 

VLV increases the dynamic instruction stream coverage by iteratively packing maximum number of scalar instructions together, even if the number is less than the number of vector lanes available. SWR employs two techniques to keep the permutations to minimum. As a result, the proposed SIMD architecture alleviate the rigidness problem of the traditional SIMD architecture and allows to generate optimized code at higher vector lengths. Moreover, the HW/SW co-designed nature of the processor provides some addition advantages. For example, since vectorization is done at runtime on the program binary, it does not require any changes in compiler, operating system or application source code. Therefore, we can target the proposed ISA without modifying anything in the software stack. 
The main contributions of this paper can be summarized as:
\begin{itemize}
\item Identifies the bottleneck in vector code generation for wider vector units.
\item Proposes a flexible SIMD architecture.
\item Proposes Variable Length Vectorization to increase the dynamic instruction stream coverage. 
\item Proposes Selective Writing to reduce the number of permutation instructions.
\end{itemize} 

This paper is an extension of our prior work~\cite{6831962} and makes the following additional contributions:
\begin{itemize}
\item Shows why both VLV and SWR are necessary and not only just either of them.
\item Shows why vector length register is not a good choice for SIMD accelerators.
\end{itemize} 

The rest of the paper is organized as follows: Section 2 provides a background on HW/SW co-designed processors. Section 3 briefly provides the motivation for the work presented in this paper and identifies key issues in efficient vector code generation for higher vector lengths. Section 4 describes the speculative dynamic vectorization algorithm. Section 5 and 6 explain the proposed SIMD ISA, Variable Length Vectorization and Selective Writing techniques. Evaluation of the proposals using a set of SPECFP2006 and Physicsbench applications is presented in Section 7. Section 8 presents related work and Section 9 concludes.

\section{Background Of HW/SW Co-designed Processors}\label{sec:Background}
A HW/SW co-designed processor is a hybrid architecture that leverages hardware/software co-design to couple a software layer to the microarchitectural design of a processor. The software layer resides between the hardware and the operating system. This software layer allows host and guest ISAs to be completely different by translating the guest ISA instructions to the host ISA dynamically. We define the host ISA as the ISA that is implemented in the hardware, whereas, guest ISA is the one for which applications are compiled. The basic idea behind these processors is to have a simple host ISA to reduce power consumption and complexity. This kind of processors\cite{Dehnert:2003:TCM:776261.776263}\cite{Ebcioglu:1997:DDC:264107.264126}\cite{Sathaye99boa:targeting} first emerged more than two decades ago. Moreover, there is a renewed interest in them in both industry and academia \cite{IntelHSCo}\cite{Lupon:2014:SHC:2541940.2541978}\cite{Brankovic:2014:WSM:2581122.2544142}
\cite{6494980}\cite{Kumar:2014:EPG:2658949.2629681}\cite{Neelakantam:2010:RSE:1736020.1736026}
\cite{6831962}\cite{6799102}. 

\begin{figure*}[t!]
\centering
\subfigure[Conventional RISC processor{\label{RISC}}]{\includegraphics[width=55mm]{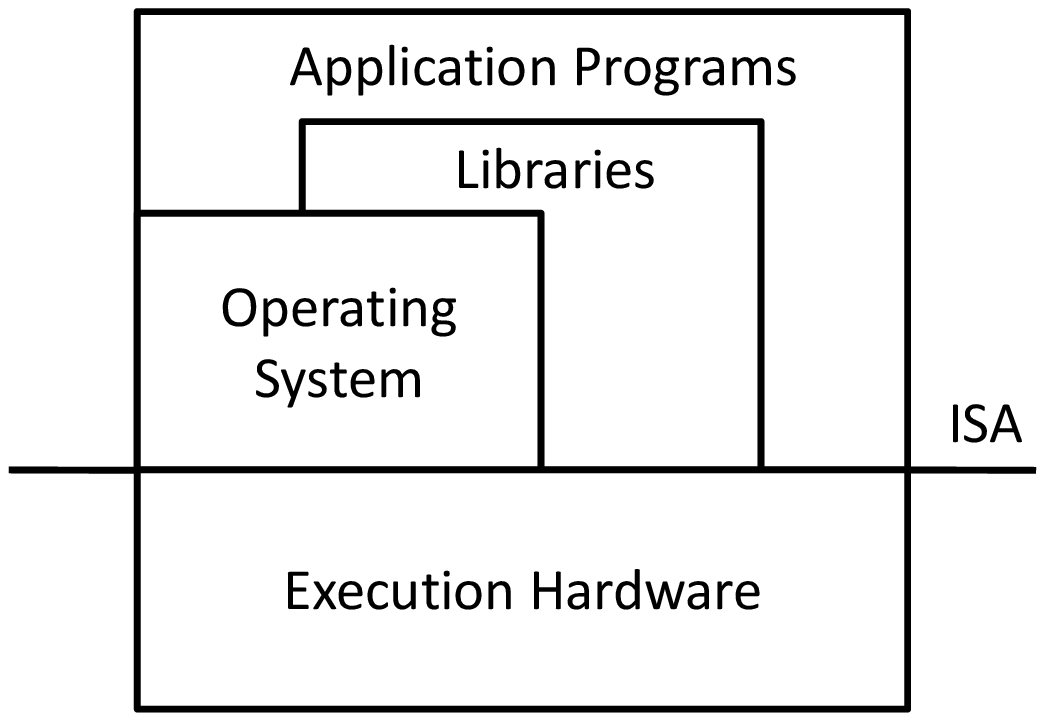}}
\subfigure[Conventional CISC processor{\label{CISC}}]{\includegraphics[width=58mm]{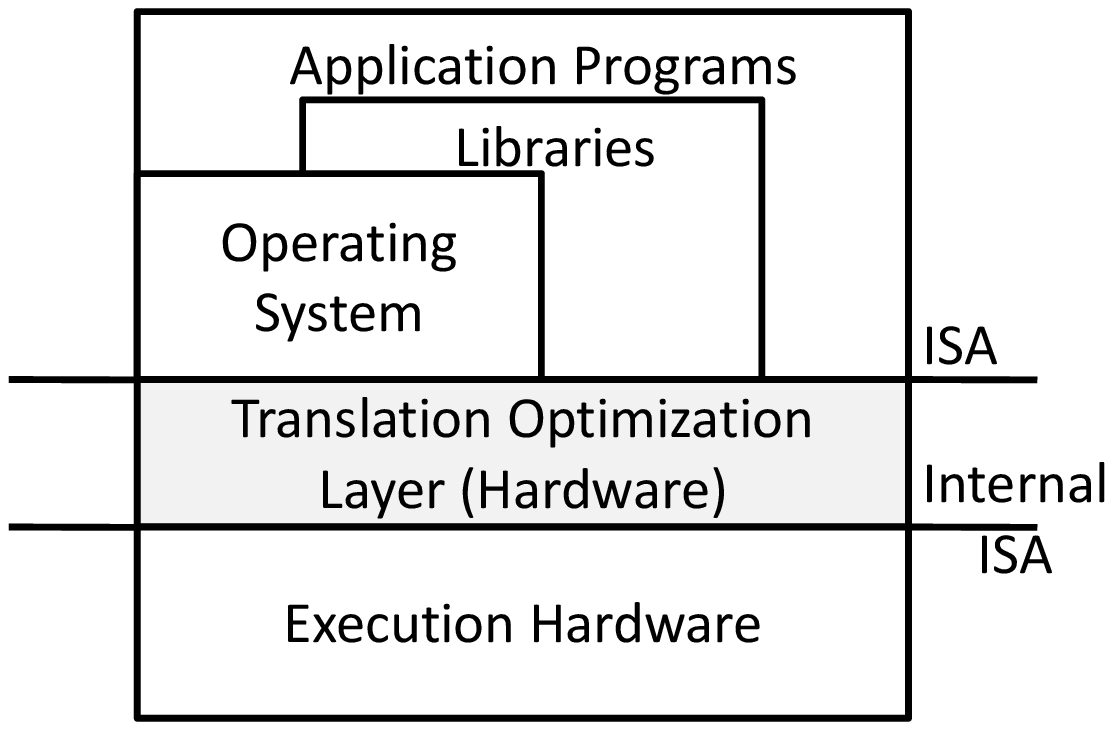}}
\subfigure[HW/SW co-designed processor{\label{HSCO}}]{\includegraphics[width=60mm]{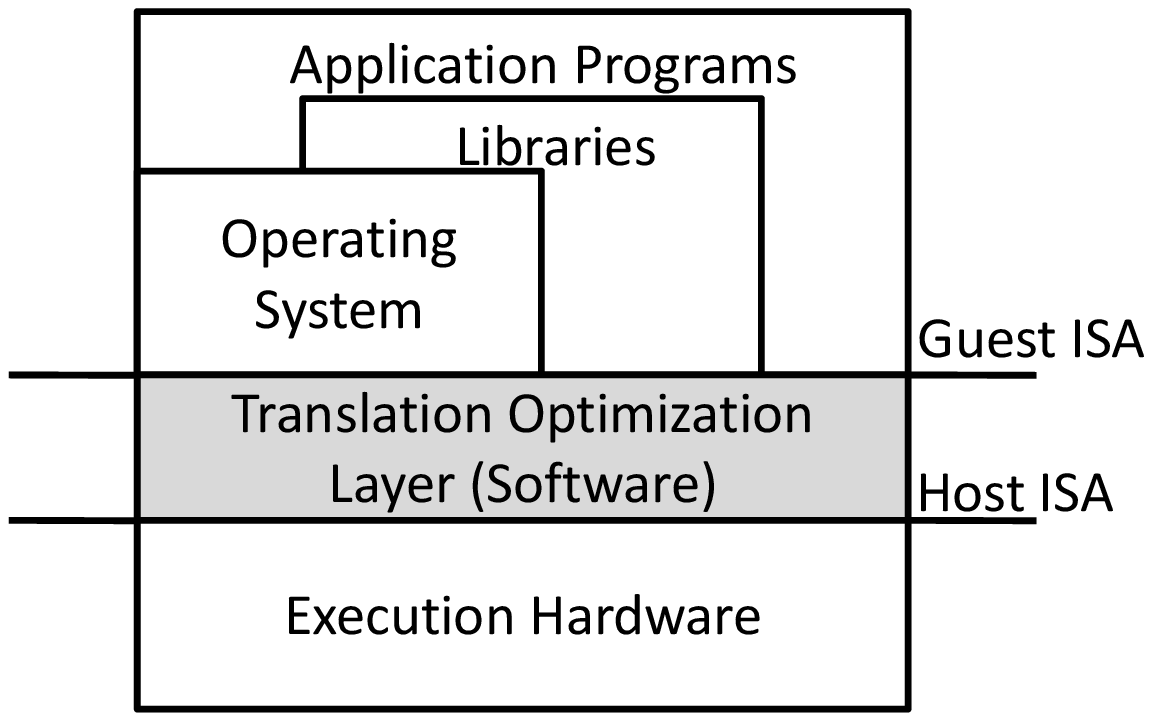}}
\caption{HW/SW interface in processors} \label{HSInterface}
\end{figure*}

These processors are specifically designed to achieve energy efficiency, design simplicity, and performance improvement. In order to achieve design simplicity, they keep the hardware simple and implement a relatively simple ISA. The simple hardware design also helps in achieving energy efficiency. Transmeta reports significant reduction in power dissipation for their HW/SW co-designed processor Crusoe compared to Intel Pentium III for a software DVD player \cite{Klaiber2000}. Their data shows that Pentium III heats up to a temperature of 105º C whereas Crusoe's maximum temperature goes only up to 48º C running the same software DVD player. Furthermore, to achieve the performance goal, HW/SW co-designed processors employ dynamic binary optimizations. 

In general, HW/SW co-designed processors implement a proprietary ISA in order to achieve design simplicity and power efficiency. Therefore, they need to apply binary translation to map the guest ISA on to the host ISA. The binary translation, in general, can be implemented in either hardware or software. Modern processors implementing CISC ISA, like x86, implement binary translation in hardware \cite{Smith:2005:VMV:1204009}. The hardware binary translator translates CISC instructions to RISC like instructions dynamically to simplify the execution pipeline implementation. However, the hardware implementation leads to significant hardware complexity and power consumption. HW/SW co-designed processors, on the other hand, implements dynamic binary translation in software which leads to energy efficiency.

Fig. \ref{RISC} shows the hardware/software interface in a conventional RISC processor where the software stack directly interacts with the hardware. Conventional CISC processors implement a RISC like ISA in hardware. As shown in Fig. \ref{CISC}, they employ a hardware dynamic binary translator to translate CISC instructions to the internal ISA instructions. The binary translation in HW/SW co-designed processors is performed by a software layer as shows Fig. \ref{HSCO}. We call this software layer as Translation Optimization Layer (TOL) in this paper.

Performing the dynamic binary translation/optimization in software layer provides several benefits over the hardware implementation. For example, the software implementation significantly reduces hardware complexity and power consumption. Furthermore, it allows to upgrade a processor in the field by introducing new optimizations in the software layer. In contrast, if TOL is implemented in hardware, adding new optimizations in the existing processor is not feasible. Additionally, software implementation of TOL significantly reduces hardware validation and verification cost and time.

\subsection{Dynamic Binary Translation/Optimization}
Translating guest ISA code to host ISA is the prime responsibility of TOL. The translation is done dynamically and, generally, in multiple phases. Usually, in the first phase, an interpreter decodes and executes guest ISA instructions sequentially. In the rest of the phases, the guest code in translated into host ISA code and stored in the code cache, after applying several dynamic optimizations, for faster execution. The number of translation phases and optimizations in each phase are implementation dependent.

\begin{figure}[t!]
\centering
\includegraphics[width=70mm]{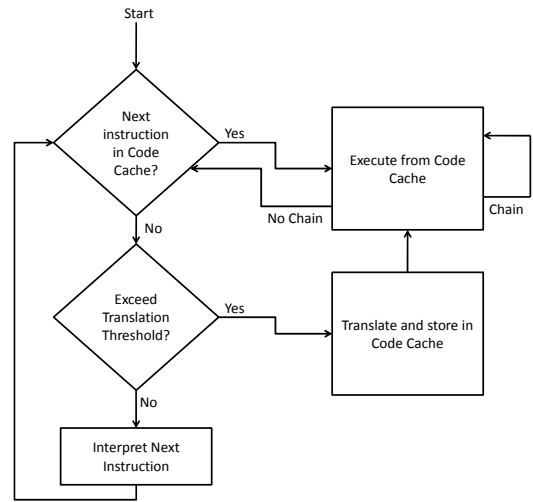}
\caption{Typical two stage TOL control flow} \label{TransOpt}
\end{figure}

Fig. \ref{TransOpt} shows a typical two stage translation/optimizations flow in a TOL. It starts by interpreting guest ISA instruction stream sequentially. While interpreting, TOL also profiles the guest code to collect information about most frequently executed code and biased branch directions. The execution frequency guides TOL to decide which guest code basic blocks to translate. When a basic block has been executed more than a predetermined number of times, TOL invokes the translator. The translator takes the guest ISA basic blocks as input, translates them to host ISA code and saves the translated code into the code cache for fast native execution. Instead of translating and optimizing each basic block in isolation, the translator uses biased branch direction information, collected during interpretation, to create bigger optimization regions, called superblocks. A superblock, generally, consists of multiple basic blocks following the biased direction of branches. Therefore, superblocks increase the scope of optimizations to multiple basic blocks and allow more aggressive optimizations. Superblocks have a single entry point that is the first instruction of the first basic block included in the superblock. However, depending on the implementation they might have multiple or a single exit point. 

Initially, the control is transferred back to TOL after executing a superblock from the code cache. Then, TOL searches the next instructions to be executed. If the next instruction is not already translated, it has to be interpreted. However, if it is already translated, TOL patches the last branch of the first superblock (the one that transferred the control back to TOL) to the beginning of the second superblock. This process is called chaining or linking. 

\subsection{Why HW/SW Co-designed Processors}
HW/SW co-designed processors provide certain features that set them apart from traditional hardware only processors. Following are the some of the reasons that motivated us to choose them for our proposals:

\textbf{Aggressive Vectorization:} Compilers inability to do accurate interprocedural pointer disambiguation and interprocedural array dependence analysis severely limits their vectorization ability\cite{Maleki:2011:EVC:2120965.2121464}. On the other hand, dynamic optimization environment in HW/SW co-designed processors avoids the need to these analysis by vectorizing speculatively\cite{6799102}. Furthermore, these processors provide efficient support to recover from speculation failures\cite{Sathaye99boa:targeting}\cite{Dehnert:2003:TCM:776261.776263}. Therefore, they enable aggressive vectorization and catch vectorization opportunities missed by conservative compiler vectorization.

\textbf{Dynamic Information:} Since the vectorization is done at runtime it benefits from the availability of the runtime information. For example, loop unroll factor can be determined at runtime through profiling for the loops where loop trip count in unknown at compile time. This is especially important for variable length vectorization where the optimal loop unroll factor varies based on logical vector length which is not always equal to the SIMD accelerator width as explained in section 5.1. 

\textbf{Decoupled vector ISA and SIMD accelerator:} HW/SW co-designed processors decouple the hardware implementation of SIMD accelerator from application visible vector ISA by means of dynamic binary translation. This enables modifications/improvements in the SIMD accelerator without affecting the application visible SIMD ISA. We leverage this fact to introduce a flexible SIMD accelerator without any modification in the application visible (guest) ISA, compiler or any other component of the software stack.

\textbf{Portable Vectorization:} Since vectorization is done by TOL at runtime, the same application binary can be executed on different SIMD accelerators. This kind of portable vectorization provides forward and backward binary compatibility.

\textbf{Legacy Code Vectorization:} Runtime vectorization in HW/SW co-designed processors also enables legacy code vectorization. Therefore, the code that was not compiled for any SIMD accelerator can also benefit from there presence.

\begin{figure}[t!]
\centering
\includegraphics[width=60mm]{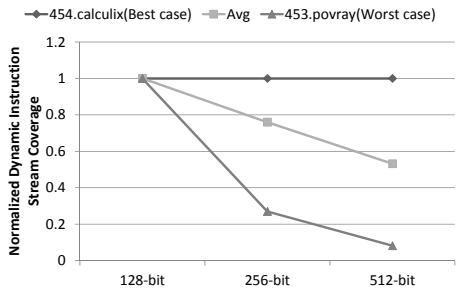}
\caption{Dynamic FP instruction stream coverage for vectorization at 128, 256 and 512-bit vector lengths} \label{Coverage}
\end{figure}

\section{Motivation}\label{sec:Motivation}
The trends in the recent past show that the vector lengths are likely to keep increasing in future microprocessors, since wider vectors provide a simple and efficient way of achieving higher FLOPS in an energy efficient manner. Intel's 256-bit AVX \cite{IntelSWDMan} and 512-bit vector length of AVX-512 \cite{IntelAVX512} and Larrabee \cite{Seiler:2008:LMX:1360612.1360617} are few examples of these trends. However, it is a challenge to generate efficient code to utilize these wider vector units. To demonstrate this fact, we vectorized floating point instructions in SPECFP2006 for three different vector lengths of 128, 256, and 512-bits using the speculative dynamic vectorization algorithm described in \cite{6799102}. Moreover, at a given vector length, all the vector instructions operate only on the maximum vector length and not on a subset of it. For example, for 512-bit vector length case, all the vector instructions operate on whole 512-bits and there is no vector instruction that operates only on 256 or 128-bits. This is inline with how the vector instructions function in the current SIMD architectures, operating on all the vector lanes and not on a subset.

Our results show that there are mainly two problems in vector code generation at higher vector lengths: reduced dynamic instruction stream coverage for vectorization and huge number of permutation instructions.

\subsection{Reduced Dynamic Instruction Stream Coverage}
We define dynamic instruction stream coverage as the number of dynamic scalar instructions vectorized. Fig. \ref{Coverage} shows the dynamic instruction stream coverage for vectorization at different vector lengths normalized to the 128-bit case.  The best, worst and average cases are shown. We divide the applications in two categories: The first category applications have maximum dynamic instruction stream coverage at all the vector lengths, like 454.calculix. On the contrary, there are applications like 444.namd where dynamic instruction steam coverage falls by 70\% at vector length of 512-bits.

The dynamic instruction stream coverage at different vector lengths depends upon the degree of data level parallelism available in the application and how this parallelism is extracted through SIMD extensions. If an application spends most of its time in loops with high trip counts, it will benefit from higher vector lengths, since the wider vector paths can be filled by unrolling the loops more number of times depending on the vector length. However, as shown by the average case of Fig. \ref{Coverage}, this is not the case for most of the applications. We see an average reduction of 25\% and 48\% in dynamic instruction stream coverage at 256-bit and 512-bit respectively. If this trend continues, the coverage is going to be even lesser at higher vector lengths.

\begin{figure}[t!]
\centering
\includegraphics[width=60mm]{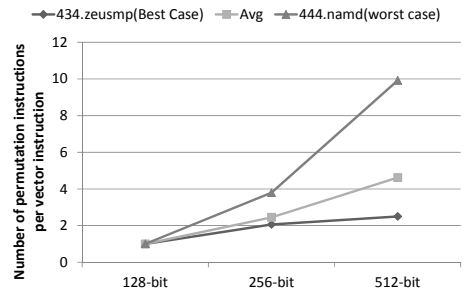}
\caption{Normalized Number of Permutation Instructions generated per vector instruction} \label{Perm}
\end{figure}

\subsection{Number of Permutation Instructions}
When the input operands of a vector instruction are not available in a single vector register or are not in the same order as required by the vector instruction, permutation instructions are needed to arrange them in the correct order. Our results show that the number of permutation instructions grows significantly with increasing vector lengths.

Fig. \ref{Perm} shows the number of permutation instructions generated per vector instruction in SPECFP2006 normalized to the 128-bit case. As the figure shows, if we generate one permutation instruction for each vector instruction at 128-bit vector length, this number goes as high as 10 at 512-bit vectors in case of 444.namd. Also, there are applications for which this number does not grow that rapidly. However, the average behavior suggests that number of permutation instructions is going to be a problem at higher vector lengths.

Both of these factors become a limitation as vector paths become wider and instead of performance improvements, it starts degrading compared to the lower vector lengths. In essence, both of these problems arise because current SIMD architectures are not flexible enough to handle these situations. The vector instructions in current SIMD architectures operate on all the vector lanes and not on a subset of it. As a result, if there are not enough independent instructions performing the same operation, compilers do not generate vector instruction. This behavior leads to reduced dynamic instruction stream coverage. Also, the scalar instructions in current SIMD architectures, such as ADDSS, MULSS etc. in Intel SSE, write their result only to lowest element of a vector register. If a vector instruction needs to read these results, they need to be packed in single register using shuffle instructions before they can be consumed by the vector instruction; thereby increasing the number of permutations. This paper investigates both the problems and proposes a flexible SIMD architecture along with Variable Length Vectorization and Selective Writing to solve the problems of reduced coverage and permutation instructions, respectively.

\section{Vectorization Algorithm}\label{sec:Vectorization Algorithm}
This section briefly discusses the baseline speculative dynamic vectorization scheme; the details of the algorithm and its evaluation can be found in \cite{6799102, VecTocs, vecPACT}. The software layer of our co-designed processor is called Translation Optimization Layer (TOL). TOL operates in three translation modes for generating host code from guest x86 code: Interpretation Mode (IM), Basic Block Translation Mode (BBM) and Superblock Translation Mode (SBM). SBM is the most aggressive translation/optimization mode and the majority (more than 90\%) of the dynamic application code is executed in this mode. Vectorization is done only in SBM, after applying several standard optimizations. 

\subsection{Pre-Vectorization Steps}
Before starting with vectorization we create a superblock, optimize them by applying standard compiler optimizations, and generate a Data Dependence Graph (DDG) as explained below:

\subsubsection{Superblock Creation}
TOL starts by interpreting guest x86 instruction stream in IM. When a basic block is executed more than a predetermined number of times, TOL switches to BBM. In this mode, the whole basic block is translated and stored in the code cache and the rest of the executions of this basic block are done from the code cache. Moreover, profiling information is gathered for all the basic blocks in BBM using software counters. This information consists of execution and edge counters. The execution counter provides the execution frequency of a basic block while the edge counters monitor the biased branch direction. Once the execution of a basic block exceeds another predetermined threshold, TOL creates a bigger optimization region, called superblock, using the branch profiling information collected during BBM. A superblock generally includes multiple basic blocks following the biased direction of branches.

Moreover, the branches inside the superblocks are converted to “asserts” so that a superblock can be treated as a single-entry, single-exit sequence of instructions. This gives the freedom to reorder and optimize instructions across multiple basic blocks. “Asserts” are similar to branches in the sense that both checks a condition. Branches determine the next instruction to be executed based on the condition; however, asserts have no such effect. If the condition is true, assert does nothing. However, if the condition evaluates to false, the assert ``fails'' and the execution is restarted from a previously saved checkpoint in IM. Furthermore, if the number of assert failures in a superblock exceeds a predetermined limit, the superblock is recreated without converting branches to “asserts”. As a result, this time the superblock has to be treated as a single-entry multiple-exit sequence of instructions. Having multiple exits in a superblock also reduces available optimization opportunities because the instructions across different exit paths cannot be reordered as freely as before.

Loop unrolling plays a major role in vectorization. Compilers unroll the loops a particular number of times to get sufficient independent instructions to fill the vector path. It is relatively simple to determine the unroll factor for loops with static trip count.  However, for the loops, where the number of iterations are not know statically, it is difficult to decide the unroll factor. The availability of dynamic application behavior in HW/SW co-designed processors allows us to detect the loop unroll factor dynamically. We profile the applications, in BBM, to collect loop iteration count for each loop. This information is used in superblock creation to decide loop unroll factor. Currently, we unroll loops with a single basic block, as the loops with no or minimum control flow are the ones which provide maximum benefits \cite{Muchnick1997}. 

\subsubsection{Pre-optimizations}
The optimizer applies several transformations on the superblock. First, x86 code is translated to an intermediate representation. Then the resulting code is transformed into a Static Single Assignment format. This transformation removes anti \& output dependences and significantly reduces the complexity of subsequent optimizations. Second, a forward pass applies a set of conventional single pass optimizations: constant folding, constant propagation, copy propagation, and common subexpression elimination. Third, a backward pass applies dead code elimination. 

After the basic optimizations, the Data Dependence Graph (DDG) is prepared. During DDG creation, we perform memory disambiguation analysis. If the analysis cannot prove that a pair of memory operations will never/always alias, it is marked as “may alias”. In case of reordering, the original memory instructions are converted to speculative memory operations. Apart from this, Redundant Load Elimination and Store Forwarding are also applied during DDG phase so that redundant memory operations are removed before vectorization. The DDG is then passed as input to the vectorizer. After vectorization, an instruction scheduler that uses a conventional list scheduling algorithm schedules the vectorized code. Afterwards, the determined schedule is used by the register allocator that implements linear scan register allocation algorithm. Finally, the optimized code is translated to the host instructions and is stored in the code cache. 

\subsection{The Vectorizer}
The vectorizer packs together a number of independent scalar instructions that perform the same operation, and replaces them with one vector instruction. The number of scalar instructions packed depends on two factors:
\begin{itemize}
\item data-types of scalar instructions
\item host vector length 
\end{itemize}

For example, for a host vector length of 128-bit, four 32-bit single-precision floating-point instructions can be packed together in a single vector instruction. Therefore, vectorization reduces dynamic instruction count and improves performance. Before describing the algorithm itself, we define a set of conditions that a pair of instructions must satisfy to be included in the same pack:
\begin{itemize}
\item The instructions must perform the same operation.
\item The instructions must be independent.
\item The instructions must not be in another pack.
\item If the instructions are load/store, they must be accessing consecutive memory locations.
\end{itemize}

Vectorization starts by marking all the instructions which are candidates for vectorization. Moreover, we mark \textit{First Load} and \textit{First Store} instructions. \textit{First Load/Store} instructions are those for which there are no other loads/stores from/to adjacently previous memory locations. For example, if there is a 64-bit load instruction $I_L$ that loads from a memory location [M] and there is no 64-bit load instruction that loads from address [M-8], we call $I_L$ \textit{First Load}.

Vectorization begins by packing consecutive stores, starting from a \textit{First Store}. The decision of starting with stores instead of loads is based on the observation that a given kind of operation always has the same number of predecessors, e.g. all the additions always have two predecessors, whereas the number of successors may vary depending on how many instructions consume the result. Consequently, following a bottom-up approach results in a more structured tree traversal than a top-down approach.

Once a pack of stores is created, their predecessors are packed, before packing other stores, if they satisfy the packing conditions. Moreover, if the last store in the pack has a next adjacent store, it is marked as \textit{First Store} so that a new pack can start from it. 

Once all the stores are packed and their predecessor/successors chains have been followed, we check for remaining load instructions that satisfy the packing conditions and pack them in the same way as stores. 

Vectorization starting from adjacent loads/stores has an obvious limitation: if a superblock does not have any consecutive loads/stores, nothing can be vectorized. To tackle this problem, after packing all loads/stores and their predecessors/successors, we check if still there are some arithmetic instructions that can be packed together. If so, we vectorize them and follow their predecessor/successor trees. This allows to partially vectorize loops with interleaved memory accesses.

While traversing the predecessor/successor chains, if we find out that the predecessors of a pack cannot be vectorized, a \textit{Pack} instruction is generated. This \textit{Pack} instruction collects the results of all the predecessors into a single vector register and feeds the current pack. Similarly, if all the successors of a pack cannot be vectorized, an \textit{Unpack} instruction is generated. This \textit{Unpack} instruction distributes the result of the pack to the scalar successor instructions. For example, in the case of loops with interleaved memory access, when we reach several load instructions while traversing the tree, we find out that they cannot be packed since they are not consecutive. Therefore, we leave them in scalar form and assemble their results using a \textit{Pack} instruction.

\section{Variable Length Vectorization}\label{sec:Variable Length Vectorization}
As shown in Fig. \ref{Coverage} in Section 3, the dynamic instruction stream coverage for vectorization reduces at higher vector lengths. We observe that the reason for this behavior lies in the way the vector instructions in SIMD architectures function. Vector instructions in the current SIMD architectures, such as ADDPS in Intel SSE, VADD in ARM Neon and VADDFP in PowerPC Altivec, operate on all the vector lanes and not on a subset of it. Due to this reason, compilers generate a vector instruction only when there are sufficient numbers of independent operations to fill the vector path. When there are not enough instructions to fill up the vector path, all the instructions are left in scalar form. This is going to be an important issue in the future microprocessors with wider vector paths and a lot of, otherwise vectorizable, code will be left unvectorized. We propose Variable Length Vectorization (VLV), a speculative dynamic iterative vectorization technique that targets a flexible SIMD architecture for optimal vectorization of data parallel applications.

VLV targets a SIMD architecture with vector instructions that can operate on all or any subset of vector lanes. Since the vector instructions can operate on any number of vector lanes, we need a way to notify the SIMD accelerator which vector lanes to enable and which ones not. We make use of mask registers for this purpose. Mask register has one bit per vector lane. The bits containing ones signify the corresponding vector lanes are to be enabled; 0 means otherwise. The mask register is included in instruction encoding in addition to the regular source and destination registers. 

An important factor to consider here is the need of masking. Masking is used to disable unused vector lanes when a vector instruction does not use all the lanes. In general, not masking the unused lanes might work well for arithmetic instructions from the functionality point of view. However, performing unnecessary operations in the unused lanes might also generate false exceptions, like divide by zero. Therefore, we would need a way to distinguish real and false exceptions. Furthermore, for memory access instructions this might result in crossing array boundaries and leading to page/segmentation faults. Also, for store instructions it would result in writing incorrect data to the memory. Moreover, the register file will contain invalid data because whole destination register will be written. As a result, we would need a way to distinguish between invalid and valid data in the register file. Mixing the architectural state and temporal values is typically not a good idea. On the other hand, masking the unused lanes helps us get rid of all these problems.

From the implementation perspective, we do not really need to have real mask registers in the hardware. Since we need to enable only consecutive lower order vector lanes, the number of lanes to be activated can directly be encoded in the instructions encoding. This also saves upon the extra instructions, otherwise, needed to write the mask in the registers. It is important to note that the traditional vector processors support variable vector length through a vector length register. It needs to be set to the desired vector length before executing vector instructions. However, it is not the optimal solutions for the processors targeting general purpose applications, where the vector length needs to be changed frequently. In this scenario, the overhead of writing the vector length register would affect the performance severely as will be shown in Section 7. Therefore, instead of having a variable vector length register we propose to have Variable Length Vectorization using masked vector instructions.

\begin{figure}[t!]
\centering
\includegraphics[width=40mm]{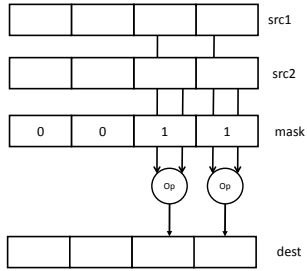}
\caption{Masked Vector Instruction Execution} \label{VLV_Exec}
\end{figure}

For the execution of a vector instruction, the hardware now reads not only the source registers but also a mask to enable only the required vector lanes. Example in Fig. \ref{VLV_Exec} shows the execution of a vector instruction that needs only two of the four vector lanes. As shown in the figure only two of the four vector lanes are activated. This is also important from the power consumption point of view, not to activate all the vector lanes for all the vector instructions.  

\subsection{Code Generation}
We modify our baseline speculative dynamic vectorization algorithm of \cite{6799102}, briefly explained in Section 4, to generate vector code with variable vector length SIMD ISA. The modified algorithm starts by vectorizing for the given maximum vector length, we call it physical vector length. Once all the possible packs for the physical vector length have been created, the vectorizer reduces the logical vector length iteratively. At lower logical vector lengths, packs are created with smaller number of scalar instructions than required to fill the vector path. The left out positions in a pack are considered as no operations.

\begin{figure}[t!]
\centering
\subfigure[Unvectorized code{\label{VLV1}}]{\includegraphics[width=70mm]{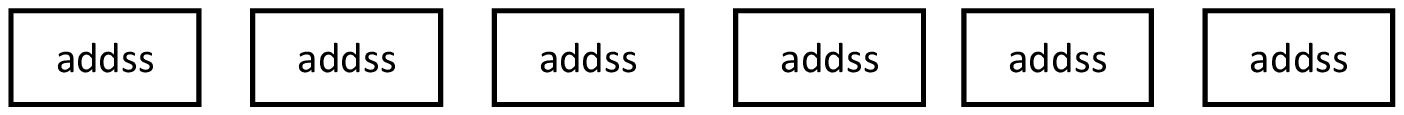}}
\subfigure[Vectorized code for fixed vector length of 128-bits{\label{VLV2}}]{\includegraphics[width=70mm]{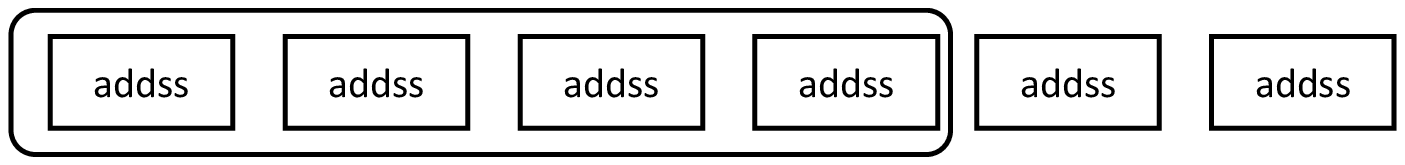}}
\subfigure[Vectorized code with variable length vectorization{\label{VLV3}}]{\includegraphics[width=70mm]{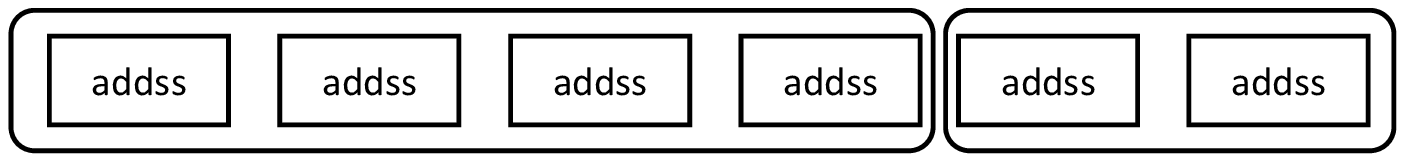}}
\caption{Variable Length Vectorization Example} \label{VLV}
\end{figure}

Fig. \ref{VLV} shows a simple vectorization example using the proposed VLV algorithm. Fig. \ref{VLV1} shows unvectorized code having six independent single-precision floating-point (32-bit) addition instructions. For a vector length of 128-bits, we can pack a maximum of four single-precision floating-point additions in a single vector addition instruction. The algorithm first packs four of the six instructions in a vector instruction and assigns a mask with all ones to this instruction, as shown in Fig. \ref{VLV2}. A mask with all ones signifies that all the vector lanes are to be enabled. 

A fixed vector length vectorization algorithm will stop at this point, since there are just two ADDSS instructions left and at least four are required to generate a vector instruction. However, VLV algorithm continues and packs the remaining two addition instructions as shown in Fig. \ref{VLV3}. Moreover, a mask register with ones only at lowest two positions is assigned to this instruction. It makes sure that only the two lower vector lanes are enabled during the execution of this vector instruction as show in Fig \ref{VLV_Exec}. 

Variable Length Vectorization helps in vectorizing the applications which have loops with lower iteration count than required by the vector length and the straight line code with fewer independent scalar operations.

VLV algorithm is fairly simple to extend to compilers for the static trip count loops, however for loops with unknown trip count at compile time it becomes tricky. For fixed vector length, compiler can vectorize such loops by unrolling them enough number of times to fill the vector path and putting a runtime check before the vectorized version to decide whether to execute it or not. However, for variable length vectorization, choosing a single unroll factor becomes difficult at compile time. The runtime information of the program behavior in HW/SW co-designed processors makes it straightforward to choose the correct unroll factor for VLV.

\section{Selective Writing}\label{sec:Selective Writing}
This section presents the proposed Selective Writing (SWR) technique to reduce the number of permutation instructions at higher vector lengths. First, we present a technique to eliminate permutation instructions completely if the result of an instruction is read only by one instruction. Then, we present another technique to reduce the number of instructions required to pack N values from N-1 to N/2, if the values to be packed are in N different registers.

\subsection{Eliminating Permutation using Selective Writing}
If the producer instructions of a vector instruction cannot be vectorized, the results of these instructions have to be packed together before feeding the vector instruction. This is due to the fact that the scalar instructions in the current SIMD architectures, such as ADDSS, MULSS etc. in Intel SIMD extensions, write their results only to the lowest element of vector registers. Whereas the vector instructions need them to be in a single vector register and in a particular order. 

\begin{figure}[h!]
\centering
\subfigure[Traditional code sequence{\label{Cswr_org}}]{\includegraphics[width=43mm]{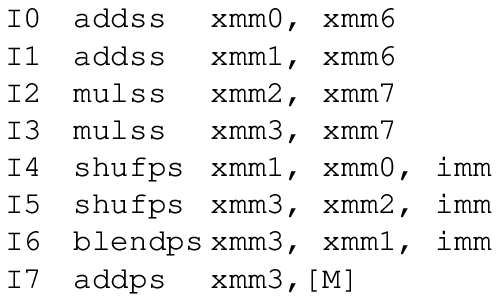}}
\subfigure[Proposed instruction sequence{\label{Cswr_new}}]{\includegraphics[width=43mm]{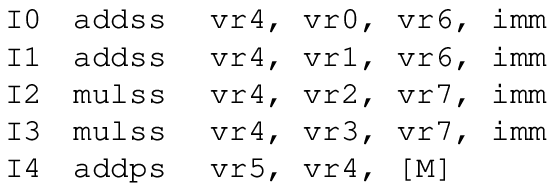}}
\caption{Packing scalar instruction results for feeding a vector instruction} \label{Cswr}
\end{figure}

Fig. \ref{Cswr_org} shows a situation where producers of I7 (I0-I3) are not vectorized and their results are packed using a permutation instruction sequence (I4-I6). As shown in the figure, I0 to I3 write their results to the lowest elements of different vector registers. Then a sequence of three instructions, I4 to I6, is used to pack these results in a single vector register xmm3, before feeding it to the vector instruction I7.

\begin{figure}[h!]
\centering
\includegraphics[width=55mm]{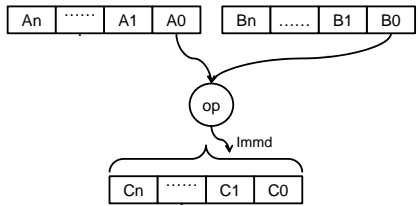}
\caption{Functionality of the proposed arithmetic scalar instructions} \label{SWR1}
\end{figure}

The scalar instructions in the proposed SIMD architecture can write their results to any element of a vector register, instead of always writing to the lowest element, thus getting rid of the permutation instructions. It is done by making the scalar instructions to selectively write to the different elements of a vector register in the order they are needed by the vector instruction, Fig. \ref{SWR1}. This way, we can avoid putting permutation instructions altogether. This kind of selective writing capability is already available in the memory access instruction set of current architectures. For example, INSERTPS in Intel SSE can be used to write a 32-bit value loaded from memory to any part of the destination register. We extend this capability to the arithmetic instruction set as well. 

In addition to carry source and destination register numbers, all scalar arithmetic instructions also carry an immediate that specifies to which element of the destination vector register the scalar result is to be written. If scalar instructions have written their results to a single vector register in the order in which they are needed by the vector instruction, the instruction sequence for packing these results is not needed anymore as shown in Fig. \ref{Cswr_new}.

The limitation of SWR scheme is that it works as long as a scalar instruction has only one consumer. In the case of more than one consumer, we would not get the maximum benefit out of SWR. However, our analysis of SPECFP2006 shows that more than 70\% of dynamic instructions have only one consumer.

\begin{figure}[t!]
\centering
\includegraphics[width=70mm]{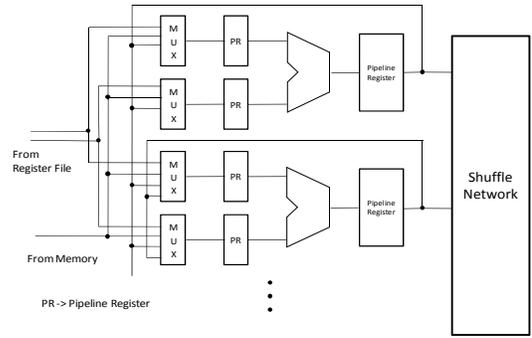}
\caption{Operand forwarding before shuffle} \label{Permhw}
\vspace{-0.1in}
\end{figure}

The proposed scalar instructions can be viewed as an arithmetic operation followed by a shuffle. However, this does not affect the latency of these instructions, since the results can be forwarded as soon as the arithmetic operation is finished. As Fig. \ref{Permhw} shows, it requires only an additional input to the multiplexers, selecting input operands of the ALUs from the output of the first vector lane (which performs scalar operations). Consequently, forwarding the results of the first vector lane to any other vector lane provides the functionality of a shuffle operation.

\subsection{Reducing Permutations to Pack N Values}
\begin{figure}[t!]
\centering
\subfigure[x86 instruction sequence{\label{Cswr1_org}}]{\includegraphics[width=43mm]{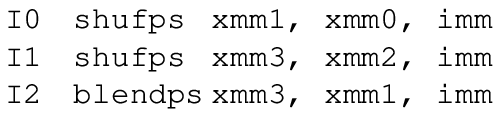}}
\subfigure[Proposed instruction sequence{\label{Cswr1_new}}]{\includegraphics[width=43mm]{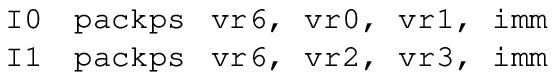}}
\caption{Instruction sequence for packing 4 values from different registers into a single register} \label{Cswr1}
\end{figure}

Current architectures provide vector instruction set where N-1 instructions are required to bring N values to a register. A typical instruction sequence to bring 4 values from different vector registers to single vector register in x86 architecture is shown in Fig. \ref{Cswr1_org}. The first two shuffle instructions bring values selected by the immediate into register xmm1 and xmm3, respectively. Then a BLENDPS instruction is used to combine the results from xmm1 and xmm3 into xmm3.

\begin{figure}[t!]
\centering
\includegraphics[width=50mm]{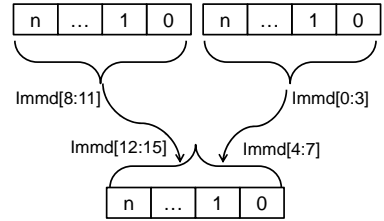}
\caption{Functionality of the proposed Pack instruction} \label{SWR2}
\end{figure}

One of the main factors that force this instruction count to be N-1 is that, these instructions write to all the elements of the destination register. If it is possible to write only the selective elements of the destination register, then this number can be brought down. In this case, the number of instructions required will depend upon the total number of different registers to be read and the number of registers that can be read by a single permutation instruction. In a case where we need to read N registers and the permutation instruction can read only two registers, we would need N/2 instructions to collect N values in a single register. If we support more number of input registers, the number of instructions required can be brought further down. Moreover, we need a mechanism to tell which elements of the source registers are to be read and which elements of the destination register are to be written.

\begin{figure*}[t!]
\centering
\includegraphics[width=0.7\textwidth, trim=0 0 0 0, clip]{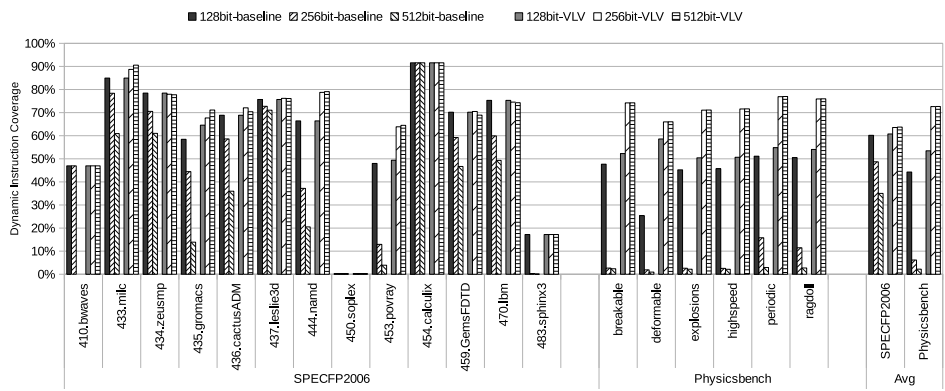}
\caption{Dynamic Instruction stream coverage at three vector lengths, baseline and with VLV} \label{Coverage_VLV}
\end{figure*}

We propose to have a permutation instruction with the functionality in Fig. \ref{SWR2}. The proposed instruction (PACKPS) has two input registers and a 16-bit immediate that tells which elements of the source and destination registers are to be accessed. The first four bits of the immediate [0:3] tells which element of the first source register is to be read and the next four bits [4:7] tell where it is to be written in the destination. Similarly, bits [8:11] tell which element of the second source register is to be written to the destination element selected by the bits [12:15]. Note that PACKPS is very similar to SHUFPS but with a bit more freedom in choosing source element for each destination element. Therefore, their latencies will be similar.

The instruction sequence for replacing x86 instruction sequence of Fig. \ref{Cswr1_org} is shown in Fig. \ref{Cswr1_new}. In this case, we are able to reduce the number of instructions required to two. For higher vector lengths, where we need to get 8 and 16 values in a register, we need just 4 and 8 instructions, respectively, instead of 7 and 15 instructions required by the original sequence. The down side of this scheme is that it requires N/2 instructions even if the values to be collected are in less than N number of registers. However, our experiments show that in SPECFP2006, on average, about 86\% and 48\% of permutations, for 256-bit and 512-bit vectors respectively, need to read N or N-1 registers to pack N values.

\section{Performance Evaluation}\label{sec:Performance Evaluation}
\subsection{Benchmarks}
To measure the success of our proposals, we use a set of applications from SPECFP2006 \cite{SPEC} and Physicsbench \cite{Yeh:2007:PAR:1250662.1250691} benchmark suites. All the SPECFP2006 benchmarks used in our experiments employ 64-bit double precision floating point data types, except 435.gromacs, whereas benchmarks in Physicsbench operate on 32-bit single precision floating point values. All the benchmarks are compiled with gcc-4.5.3 with  “-O3 -fomit-frame-pointer -ffast-math -mfpmath=sse -msse3” flags.

For SPECFP2006 we instrument the benchmarks, using PIN \cite{Luk:2005:PBC:1065010.1065034}, to find the most frequently executing routines. Then we simulate one billion instructions starting from these routines. The benchmarks in Physicsbench are executed till completion.

\begin{figure*}[t!]
\centering
\includegraphics[width=0.9\textwidth]{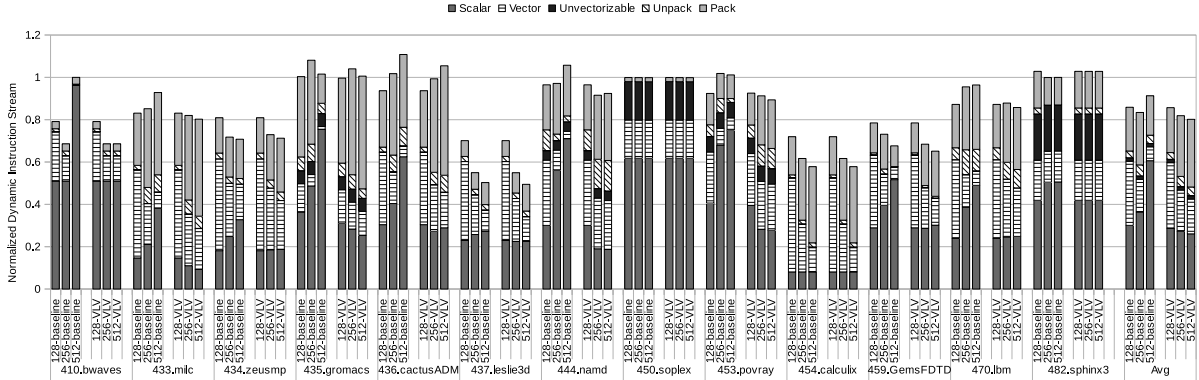}
\caption{Dynamic Instruction stream distribution for SPECFP2006: 128, 256 and 512-bit vector lengths without and with VLV} \label{VLV_only}
\end{figure*}

\subsection{Experimental Framework}
To evaluate our proposals, we use DARCO \cite{Pavlou2011, darco}, which is an infrastructure for evaluating HW/SW co-designed virtual machines. DARCO executes guest x86 binary on a PowerPC-like RISC host architecture. Since DARCO emulates floating point code in software, we extended the infrastructure to add floating point scalar and vector operations. We implemented the dynamic vectorization algorithm in the TOL to provide vectorization support.

For our experiments, we extended the host architecture to supports vector sizes of 128, 256 and 512-bits. Moreover, we consider only floating point operations for vectorization (because most SIMD optimizations tend to focus on them) and no integer operation is vectorized. Therefore, we show only the floating point instructions in the results presented. 

\subsection{Dynamic Instruction Stream Coverage}

Fig. \ref{Coverage_VLV} shows the dynamic instruction stream coverage for three vector lengths first without and then with Variable Length Vectorization (VLV). We will have maximum coverage when the number of instructions required to create a pack is minimum, i.e. two instructions. At 128-bit vector length the maximum number of 64-bit double precision operations that can be packed together is two. Therefore, 128-bit vector length provides maximum coverage, even without VLV, for double precision operations. Since all the SPECFP2006 benchmarks primarily operate on double precision floating point variables, they have maximum coverage at 128-bits as shown in Fig. \ref{Coverage_VLV}. For single precision floating point variables, Variable Length Vectorization helps increasing coverage even at 128-bit vector length, as is evident from the figure, for Physicsbench benchmark suite and 435.gromacs.

For the vector lengths of 256-bit and 512-bits, the benchmarks can be divided into two categories. First, the benchmarks like 454.calculix have maximum, or close to maximum, dynamic instruction stream coverage at higher vector lengths also. The hottest loops of these benchmarks have enough iterations to fill the wider vector paths. Second, the benchmarks like 436.cactusADM, 444.namd, and Physicsbench show drastic reduction in coverage as vector length increases, due to the lack of independent instructions to fill the wider paths. These benchmarks either have loops with fewer iterations or with complex control flow. For example, the hottest loops in 410.bwave iterate four times, therefore, for 256-bit vector length it has the maximum coverage but for 512-bit, it drops down to zero. Benchmarks in Physicsbench have loops with complex control flow and cannot be unrolled. Moreover, number of independent instruction in individual superblocks is not enough to fill the vector path. Thus, the dynamic instruction stream coverage reduces severely.  Using VLV, we bring the coverage for these benchmarks also to the maximum as shown in the Fig. \ref{Coverage_VLV}.

\begin{figure*}[t!]
\centering
\includegraphics[width=0.7\textwidth]{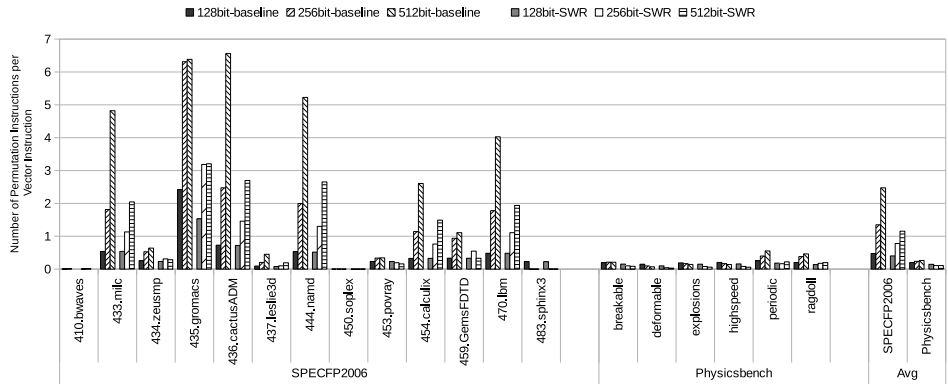}
\caption{Number of Permutation Instructions per vector instruction, baseline and with SWR} \label{Perm_SWR}
\end{figure*}

\begin{figure*}[t!]
\centering
\includegraphics[width=0.9\textwidth]{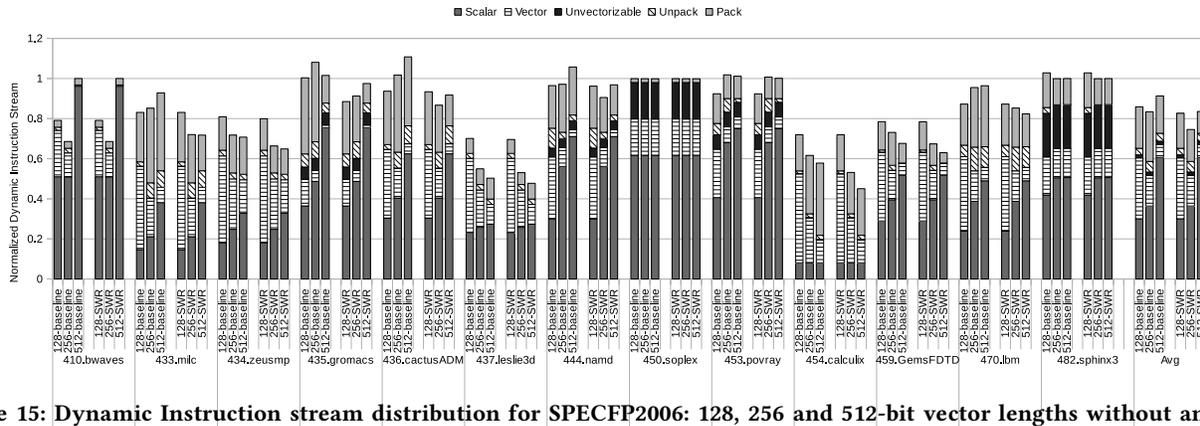}
\caption{Dynamic Instruction stream distribution for SPECFP2006: 128, 256 and 512-bit vector lengths without and with SWR} \label{SWR_only}
\end{figure*}

\subsection{Dynamic Instruction Stream Distribution with VLV}
This section shows that even though VLV increases the dynamic instructions stream coverage, by itself it does not provide much benefit in terms of overall dynamic instruction reduction because of a corresponding increase in permutations. Fig. \ref{VLV_only} presents dynamic instruction stream distribution for SPECFP2006 for 128, 256 and 512-bit vector lengths first without VLV (called baseline in the figure) and then with VLV. The results shown are normalized to no vectorization case. The dynamic instruction stream is divided into: Scalar and Vector instructions, Pack/Unpack instructions (as described in Section 4.2), and unvectorizable instructions (e.g. we do not vectorize conversions).

On average, the number of scalar instructions increases with increase in vector length without VLV as shown by the 128, 256 and 512-bit baseline case. Scalar instructions constitute 31\% of overall dynamic instruction stream for SPECFP2006 at 128-bit vector length without VLV. However this number increases to 41\% and 52\% at 256 and 512-bit without VLV. It is because of this increase in scalar instructions (or the corresponding decrease in dynamic instruction stream coverage) that we do not get any reduction in overall dynamic instruction stream at higher vector lengths. VLV, on the other hand, reduces the scalar instructions in the dynamic instruction stream by extracting additional vectorization opportunities. As shown in Fig. \ref{VLV_only}, VLV brings down the scalar instructions to 28\% from 41\% and 52\% at 256 and 512-bit vector lengths.

Even though VLV increases the dynamic instructions vectorized, the overall reduction in dynamic instructions stream is only marginal as is evident from Fig. \ref{VLV_only}. It is the result of the fact that the increased number of vectorized instructions comes at the cost of an increase in the permutations. Therefore, we need a way to keep the permutation instructions to a minimum. We use Selective Writing (SWR) as a means to that and evaluate it next.

For Physicsbench, VLV by itself is able to provide significant dynamic instruction stream reduction with minimal increase in permutations. Therefore, we do not show results for it.

\subsection{Permutation Reduction}

Fig. \ref{Perm_SWR} shows the number of permutation instructions per vector instruction required at three vector lengths without and with Selective Writing (SWR). Again, we have the same two categories of benchmarks as for the dynamic instruction stream coverage. Benchmarks like 434.zeusmp, 459.GemsFDTD, and Physicsbench have, essentially, the same amount of permutation instructions across all the vector lengths. Packing the instructions from the different iterations of unrolled loops avoids generation of permutation instructions in the case of 434.zeusmp and 459.GemsFDTD. Physicsbench, however, has really less number of permutations since we fail to vectorize anything. On the contrary, 433.milc, 436.cactusADM and 444.namd show an increase in the permutation instructions at higher vector lengths. Complex control flow and fewer loop iterations forces us to vectorize straight line code which require higher number of permutation instructions. SWR helps in eliminating significant number of permutation instructions for these benchmarks.

Another point to notice in Fig. \ref{Perm_SWR} is that for 128-bit vector length there is negligible reduction in permutation instructions. This is because we need to pack two double precision values in a 128-bit register and for N=2, N/2 and N-1 are same. Therefore, we do not get much benefit. However, on average we reduce the number of permutation instruction required to half.

\subsection{Dynamic Instruction Stream Distribution with SWR}
This section shows that even though SWR is effective in keeping the permutation instructions to a minimum, it also by itself is unable to provide significant overall dynamic instruction reduction. Fig. \ref{SWR_only} present dynamic instruction stream distribution for SPECFP2006 for 128, 256 and 512-bit vector lengths first without SWR (called baseline in the figure) and then with SWR. The results shown are also normalized to no vectorization case. The dynamic instruction stream is again divided into: Scalar and Vector instructions, Pack/Unpack instructions and unvectorizable instructions.

SWR achieves significant permutation reduction as shown in Fig. \ref{SWR_only} especially for 433.milc, 436.cactusADM and 470.lbm benchmarks. For other benchmarks like 410.bwaves, 434.zeusmp, 437.leslie3d etc. permutation instructions are not significant either because of small number of vectorized instructions due to less coverage or because the benchmarks have enough parallelism at higher vector lengths also. Even though SWR is effective in keeping the permutations to a minimum it cannot provide significant dynamic instruction reduction if the vectorizer is not able to vectorize most of the code as shown in Fig. \ref{SWR_only}. 

Therefore, none of VLV and SWR by itself is able to achieve significant dynamic instruction stream reductions at higher vector lengths. However, when combined together, they do reduce the dynamic instruction stream substantially as shown in the next section. 

\subsection{Putting Everything Together}
\begin{figure*}[t!]
\centering
\includegraphics[width=0.7\textwidth]{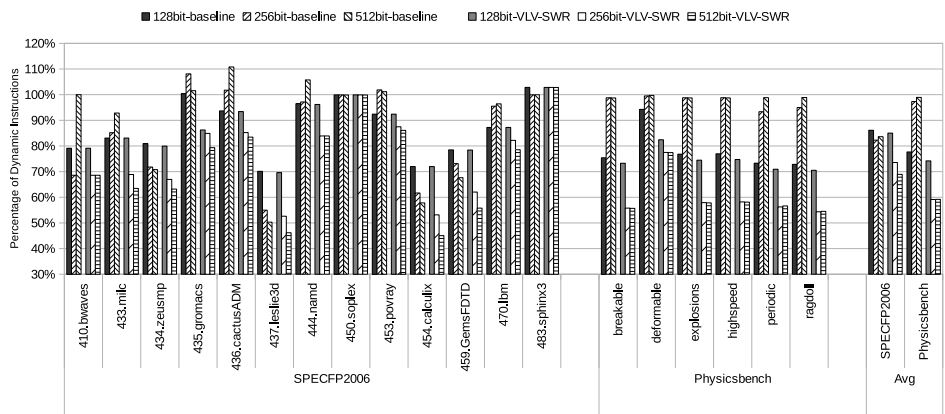}
\caption{Dynamic Instruction Percentage after baseline and VLV-SWR vectorizations} \label{DynInsn_VLV_SWR}
\end{figure*}

Fig. \ref{DynInsn_VLV_SWR} shows the percentage of dynamic instructions after vectorization without and with VLV-SWR. As shown in this figure, after applying both the optimizations all the applications perform better as vector length is increased. Applications like 433.milc, 436.cactusADM, 470.lbm, and Physicsbench which were earlier getting worse with increase in the vector length, compared to 128-bit vector length; now perform better. On average, VLV-SWR help eliminating 9\% and 16\% more dynamic instructions compared to the baseline vectorization, at 256-bit and 512-bit vector lengths respectively, for SPECFP2006. Overall, vectorization with VLV-SWR reduce unvectorized dynamic instruction stream by 15\%, 27\%, and 31\% for 128-bit, 256-bit, and 512-bit vector lengths respectively. For Physicsbench, we eliminate 40\% more instructions compared to baseline vectorization and unvectorized code, at 256-bit, and 512-bit vector lengths with VLV- SWR. Baseline vectorization does not find any vectorization opportunity at higher vector lengths for Physicsbench.

As Fig. \ref{DynInsn_VLV_SWR} shows, the percentage of reduced instructions is same for 256-bit and 512-bit vector lengths in case of Physicsbench and 410.bwaves. The lack of availability of independent instructions at 512-bit vector length forces VLV to vectorize the code the same way as for 256-bit vector length. However, important point to notice is that we still have more instruction reduction than 128-bit case, which was not possible without VLV.

\subsection{Vector Length Register vs VLV}
\begin{figure}[t!]
\centering
\includegraphics[width=70mm]{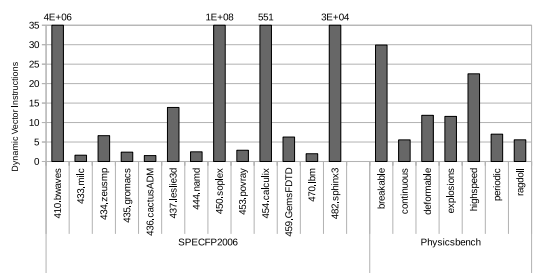}
\caption{Average number of consecutive dynamic vector instructions with same vector length in a 512-bit wide vector unit with VLV-SWR} \label{VLR}
\end{figure}

Traditional vector processors used a special register, called vector length register, to choose the number of vector lanes to be enabled. This register needs to be written every time a vector instruction needs different number of lanes than the vector instruction immediately preceding it. This section shows why vector length register is not an optimal solution in SIMD accelerators for dynamically varying the logical vector length. Fig. \ref{VLR} shows the average number of dynamic vector instructions executed before a vector instruction requiring a different number of vector lanes is encountered. In other words, the figure shows how frequently the vector length register would need to be written had we used it instead of the proposed VLV.

As the figure shows, a hypothetical vector length register would need to be written very frequently for most of the benchmarks. For example, for 433.milc, 436.cactusADM and 470.lbm it would be written after executing only two vector instructions. Although, there are few benchmarks like 410.bwaves, 454.calculix and 482.sphinx3 where the writes to the vector length register are quite rare however, for the majority of the benchmarks it would need to be written very frequently. The vector processors could use vector length register because they specifically targeted heavily data parallel applications. 

The extra instructions to write the vector length register would severely affect the performance benefits of vector execution. Therefore, VLV chooses to encode the number of vector lanes to be enabled in the instruction encoding rather than using a vector length register. 

\subsection{Performance}

We model a simple in-order processor, in congruence with the simple hardware design philosophy of the co-designed processors, with issue width of two. Microarchitectural parameters are shown in Table 1.

\begin{table}[t]
\caption{Processor Microarchitectural Parameters}
\label{tableratio}
\small
\begin{center}
\begin{tabular}{|p{2.5cm} |p{5cm} |}
\hline
\textbf{Parameter} & \textbf{Value}\\
\hline
L1 I-cache & 64KB, 4-way set associative, 64-byte line, 1 cycle hit, LRU \\
\hline
L1 D-cache & 64KB, 4-way set associative, 64-byte line, 1 cycle hit, LRU \\
\hline
Unified L2 cache & 512KB, 8-way set associative,  64-byte line, 6 cycle hit, LRU \\
\hline
Scalar Functional Units (latency) & 2 simple int(1), 2 int mul/div (3/10) 2 simple FP(2), 2 FP mul/div (4/20) \\
\hline
Vector Functional Units (latency) & 1 simple int(1), 1 int mul/div (3/10) 1 simple FP(2), 1 FP mul/div (4/20) \\
\hline
Registers & 128-Integer, 128-Vector, 32-FP \\
\hline
Memory Lat & 128 Cycles \\
\hline
\end{tabular}

\end{center}
\end{table}

\begin{figure*}[t!]
\centering
\includegraphics[width=0.8\textwidth]{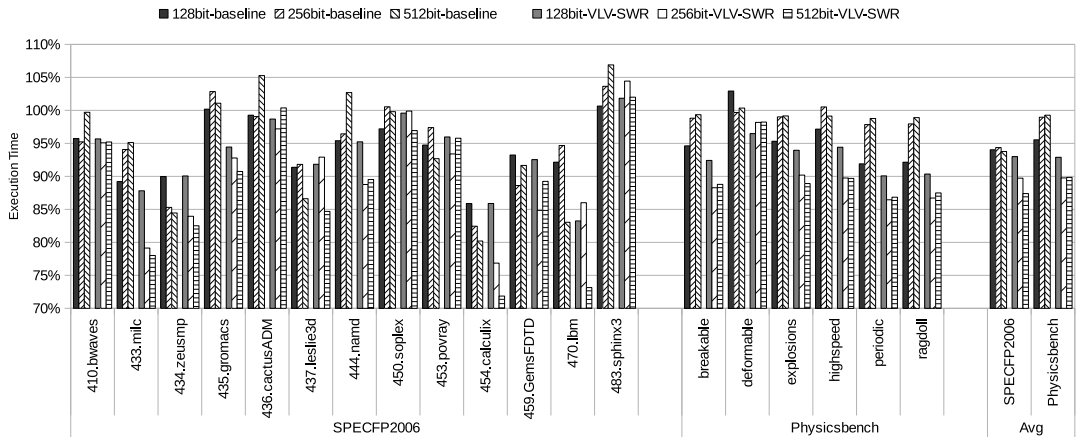}
\caption{Execution time for baseline and VLV-SWR vectorizations normalized to unvectorized code execution time} \label{perf}
\end{figure*}

Fig. \ref{perf} shows the percentage of execution time, at three vector lengths, after vectorization without and with VLV-SWR. On average VLV-SWR provide 5\% and 7\% speed up over the baseline vectorization and 10\% and 13\% over the unvectorized code, for vector length of 256-bit and 512-bit respectively, for SPECFP2006. Similarly, for Physicsbench, we get a speed up of 10\% for with VLV-SWR over unvectorized and baseline vectorization.

There are several interesting points to note in Fig. \ref{perf}. First, even though we have higher dynamic instruction elimination, e.g. 31\% for SPECFP2006, the speed up we get is smaller, 13\% for SPECFP2006 at 512-bit vector length. This is because only 39\% of dynamic instructions are floating point in SPECFP2006, which reduces the overall performance. Second, dynamic instruction reduction is more for Physicsbench, 40\% compared to 31\% of SPECFP2006 for 512-bit vector length; SPECFP2006 shows more speed up, 13\% compared to 10\% of Physicsbench for 512-bit vector length. This is due to the fact that Physicsbench has higher percentage of integer instructions than SPECFP2006.

\section{Related Work}\label{sec:Related Work}
Masked operations have been used in the past for vectorization of code with control flow. However, we use them in the absence of control flow to increase dynamic instructions stream coverage. J. Smith et al. \cite{Smith:2000:VIS:339647.339693} proposed masked operations as a means of adding support for conditional operations in vector instruction set. J. Shin et al. \cite{Shin:2005:SPP:1048922.1048983} incorporated masked operations to vectorize loops with conditional flow in Superword Level Parallelism approach. Larrabee also uses masked instructions to map scalar if-then-else control structure to the vector processing unit. All of these proposals execute both if and else clauses and select the correct results based on the values in the mask registers. Our proposal, on the other hand, uses masked operations to increase the dynamic instruction stream coverage when there not enough instruction to fill the wider vector paths.
 
Significant amount of work has been done on the optimal generation of permutation instructions. However, previous work does not show effect of permutations at increasing vector lengths. A. Kudriavtsev et al. \cite{Kudriavtsev:2005:GPS:1065910.1065931} show the relationship between operation grouping and permutation generation. They show the ordering of individual operations in SIMD instructions affect the number of permutation instructions required. G. Ren et al. \cite{Ren:2006:ODP:1133981.1133996} presented an algorithm that converts all the permutations to a generic form. Then, permutations are propagated across the statement and redundant permutations are eliminated. These solutions focus on reducing the number of permutations required, whereas our solution reduces the number of instructions for each permutation.  L. Huang et al. \cite{5416631} proposed a method to reduce the number of instruction for one permutation. Their system has a Permutation Vector Register File which provides implicit permutation capabilities. However, the permutation pattern is to be saved beforehand in a permutation register. Moreover, only the values from two consecutive registers can be permutated.

The proposal by M. Woh et al. \cite{Woh:2009:AAA:1555754.1555773} for supporting multiple SIMD widths is the closest to our proposal of Variable Length Vectorization. They proposed a configurable SIMD datapath that can be configured to process wide vectors or multiple narrow vectors. Unfortunately, details of their vectorization algorithm for vectorization for multiple vector lengths are not provided. 

Speculative Dynamic Vectorization, in itself, is not a much extended topic in literature. There have only been a few proposals like Speculative Dynamic Vectorization \cite{1003585}, Dynamic Vectorization in Trace Processors \cite{Vajapeyam99} and Liquid SIMD \cite{4147662}. None of them is in the context of HW/SW co-designed processors. A. Pajuelo et al. \cite{1003585} proposed to speculatively vectorize the instruction stream in the hardware for superscalar architectures. Their scheme prefetches data into the vector registers and speculatively manipulates it through arithmetic instructions. S. Vajapeyam et al. \cite{Vajapeyam99} builds a large logical instruction window and converts repetitive dynamic instructions from different iterations of a loop into vector form. The whole loop is vectorized if all iterations of the loop have the same control flow. Liquid SIMD \cite{4147662} decouples the SIMD accelerator implementation from the instruction set of the processor by compiler support and a hardware based dynamic translator. Compiler passes hints to dynamic translator, which can then retarget the vector code for different SIMD accelerators. Selective devectorization~\cite{devec, Kumar:2014:EPG:2658949.2629681} has also been explored to reduce the energy consumption of SIMD accelerators by keeping them power gated for longer intervals.

\section{Conclusion}\label{sec:Conclusion}
In this paper, we showed that widening the SIMD accelerators does not improve the performance for all the applications. We discovered two main problems hurting the performance of naturally low vector length applications for wider SIMD units: Reduced dynamic instruction stream coverage and large number of permutation instructions. 

We proposed a flexible SIMD architecture that allows the vector instructions to operate on variable number of lanes. Additionally, the scalar instructions can selectively write to any element of the vector register, thus avoiding permutations. We also proposed Variable Length Vectorization and Selective Writing techniques to target the flexibility of the proposed SIMD architecture. Variable Length Vectorization vectorizes the code even though it is not possible to fill the wider vector path. Selective Writing allows to write to any particular element of vector registers, thus reduces permutations. Our experimental results show an average dynamic instruction elimination of 31\% and 40\% and an average speed up of 13\% and 10\% for SPECFP2006 and Physicsbench respectively, for 512-bit vector length, over the scalar baseline code.

\bibliographystyle{IEEEtranS}
\bibliography{references}

\end{document}